\documentclass[preprint,12pt,a4paper]{elsarticle}

\usepackage{amssymb}
\usepackage{hyperref}
\usepackage{verbatim}
\usepackage{booktabs}
\usepackage{listings}
\usepackage{xcolor}
\usepackage{tikz}
\usetikzlibrary{positioning, backgrounds, arrows.meta}
\usepackage{multirow}

\definecolor{codegreen}{rgb}{0,0.6,0}
\definecolor{codegray}{rgb}{0.5,0.5,0.5}
\definecolor{codepurple}{rgb}{0.58,0,0.82}
\definecolor{backcolour}{rgb}{0.95,0.95,0.92}

\journal{SoftwareX}

\begin{document}
\renewcommand{\labelenumii}{\arabic{enumi}.\arabic{enumii}}

\begin{frontmatter}

\title{JAX-AMG: A GPU-Accelerated Differentiable Sparse Linear Solver Library for JAX}

\author[cornellMAE,ndAME]{Yi Liu}
\author[cornellMAE]{Xiantao Fan}
\author[cornellMAE,ndAME]{Jian-Xun Wang\corref{corxh}}

\address[cornellMAE]{Sibley School of Mechanical and Aerospace Engineering, Cornell University, Ithaca, NY, USA}
\address[ndAME]{Department of Aerospace and Mechanical Engineering, University of Notre Dame, Notre Dame, IN, USA}

\cortext[corxh]{Corresponding author. Tel: +1 540 315 6512}
\ead{jw2837@cornell.edu}

\begin{abstract}

Sparse linear systems from PDE discretizations are central to scientific computing, yet no existing JAX-ecosystem solver simultaneously provides GPU-accelerated algebraic multigrid (AMG), automatic differentiation (AD), and distributed multi-GPU execution. JAX-AMG fills this gap by wrapping the NVIDIA AmgX solver suite as a native JAX primitive, exposing AMG and Krylov methods with configurable preconditioners through a unified interface compatible with JIT compilation, reverse-mode AD via adjoint methods, and MPI-based distributed execution. Solver caching amortizes setup costs across repeated solves, making JAX-AMG practical for PDE-constrained optimization and inverse problems. The result is a robust, scalable sparse linear algebra layer that integrates seamlessly into differentiable simulation and scientific machine learning pipelines.

\end{abstract}


\end{frontmatter}

\section*{Metadata}

\begin{table}[!h]
\begin{tabular}{|l|p{6.5cm}|p{6.5cm}|}
\hline
\textbf{Nr.} & \textbf{Code metadata description} & \textbf{Metadata} \\
\hline
C1 & Current code version & v0.1.4 \\
\hline
C2 & Permanent link to code/repository used for this code version & \url{https://github.com/jx-wang-s-group/JAX-AMG/tree/v0.1.4} \\
\hline
C3 & Permanent link to Reproducible Capsule &  \\
\hline
C4 & Legal Code License & Apache License 2.0 \\
\hline
C5 & Code versioning system used & git \\
\hline
C6 & Software code languages, tools, and services used & Python, C++, CUDA, MPI \\
\hline
C7 & Compilation requirements, operating environments \& dependencies & Python 3.10+, CUDA Toolkit 12+, NVIDIA AmgX 2.5+, JAX 0.5.0+ (with CUDA support), MPI library (with mpi4py and mpi4jax) \\
\hline
C8 & If available Link to developer documentation/manual & \url{https://jx-wang-s-group.github.io/JAX-AMG/} \\
\hline
C9 & Support email for questions & \url{jw2837@cornell.edu} \\
\hline
\end{tabular}
\caption{Code metadata}
\label{codeMetadata}
\end{table}

\section{Motivation and significance}

Sparse linear systems arising from the discretization of partial differential
equations (PDEs) are ubiquitous in computational science and engineering, with critical applications across fluid mechanics \cite{ferziger2020computational}, heat transfer \cite{ozicsik2017finite}, structural mechanics \cite{hughes2000finite}, electromagnetics \cite{jin2014finite}, and many other disciplines. Increasingly, the solvers for these systems are embedded within gradient-based optimization and machine learning pipelines that require end-to-end differentiability. When a PDE solver is part of a loss function, for instance in design optimization, parameter inference, or hybrid neural--physics model training, the underlying sparse linear solver must support automatic differentiation (AD) so that gradients can propagate backward through the linear solve to update design parameters, operator coefficients, or network weights. This places new demands on sparse linear solver software, which now needs to be not only fast and robust but also compatible with modern AD workflows.

The need to differentiate through PDE solvers is not new in spirit. PDE-constrained optimization is a mature field that combines numerical simulation, sensitivity analysis, and optimization to solve design, control, and inverse problems governed by PDEs. Classical approaches have relied on adjoint-based derivative computations~\cite{hinze2009optimization}. In fluid dynamics, for example, adjoint design methods grew out of optimal-control ideas, including early work on optimum design by Pironneau~\cite{pironneau1974optimum} and pioneering adjoint-based design optimization by Jameson and collaborators~\cite{jameson1988aerodynamic,jameson1994control,jameson1998optimum}. These methods were later extended and adopted more widely in aerodynamic optimization~\cite{elliott1997practical,anderson1999aerodynamic,giles2000introduction}. Open-source frameworks such as SU2~\cite{economon2016su2} for computational fluid dynamics (CFD) and dolfin-adjoint~\cite{farrell2013automated} for finite-element PDE workflows have since made adjoint-based PDE-constrained optimization more practical in scientific computing.

Beyond these PDE-specific adjoint frameworks, general-purpose differentiable programming offers a more versatile route in modern scientific computing: building end-to-end differentiable workflows by composing discretized solvers, learnable neural networks, data assimilation, and uncertainty quantification within a single programming system. This hybrid neural--physics paradigm  enables inverse problems, parameter inference, and optimal control through physics-as-architecture designs where the governing equations are embedded directly into the model architecture rather than  enforced through loss functions~\cite{fan2024differentiable,fan2025neural,akhare2025hybridndiff}. JAX has emerged as a prominent ecosystem in this direction, combining array programming, AD, and just-in-time (JIT) compilation~\cite{jax2018github}. However, the bottleneck for such hybrid architectures is not the neural 
network, but the robust solution of large-scale sparse linear systems arising from the discretized PDEs within the backward pass. A prominent example arises in CFD when solving the pressure Poisson equation for the incompressible Navier--Stokes equations on complex computational meshes. The resulting linear systems are frequently ill-conditioned, and standard iterative solvers such as the conjugate gradient (CG) method or biconjugate gradient stabilized (BiCGSTAB) method often converge slowly or fail entirely on such systems. Algebraic multigrid (AMG) is one of the most effective approaches for these problems, using a hierarchy of progressively coarser representations to efficiently resolve errors across multiple scales \cite{stuben2001review}. Mature, high-performance AMG implementations are widely deployed across large-scale engineering and scientific computing, including hypre's BoomerAMG~\cite{falgout2002hypre,yang2002boomeramg}, the PETSc toolkit with its native GAMG solver~\cite{petsc1997}, the Trilinos ML/MueLu multigrid packages~\cite{gee2006ml,muelu2019}, and the header-only AMGCL library~\cite{demidov2019amgcl}, all of which support MPI and, in several cases, GPU execution. These libraries are implemented in compiled languages (C/C++/Fortran) and are not integrated with AD or the JAX transformation system, and thus cannot be embedded directly into the end-to-end differentiable, GPU-native pipelines targeted here. Consequently, no high-performance AMG solver in the JAX ecosystem integrated both automatic differentiation and GPU-accelerated distributed execution.

The current Python software landscape addresses isolated parts of this problem but fails to provide a unified solution. For instance, {PyAMG} \cite{pyamg2023} offers a mature AMG implementation but operates exclusively on CPUs. The {pyamgx} package~\cite{pyamgx} provides a Pythonic interface to the GPU-accelerated AmgX backend API; however, it is restricted to single-GPU execution and lacks both AD and tracing capabilities. Similarly, {petsc4py}~\cite{petsc4py2011}, which wraps the PETSc library~\cite{petsc1997} and supports GPU execution, AMG, and MPI, also lacks both AD and JIT compatibility, preventing its use in  end-to-end differentiable pipelines. Conversely, differentiable sparse linear solvers within the JAX ecosystem, whether provided directly by JAX or by libraries like Lineax~\cite{lineax2023}, currently lack support for AMG, leaving their iterative methods without the robust preconditioning required for ill-conditioned systems. A contemporaneous effort, AMJax~\cite{amjax2026}, addresses this missing AMG capability from a complementary PyAMG/JAX direction: it builds the multigrid hierarchy on CPU with PyAMG and performs AMG solves in JAX. This design is lightweight and portable across JAX-supported backends after setup, but it does not target high-performance GPU-native AMG or distributed multi-GPU execution (a detailed performance comparison between JAX-AMG and AMJax is provided in Section~\ref{sec:performance}). Similar limitations also arise beyond the JAX ecosystem: GPU-native differentiable frameworks such as NVIDIA Warp~\cite{warp2022} likewise provide JIT compilation and AD, but expose only Krylov solvers with basic preconditioning rather than AMG, and are oriented toward implementing custom differentiable simulation kernels rather than serving as drop-in solvers within JAX-based pipelines. Together, these limitations motivate JAX-AMG as a bridge between high-performance GPU sparse solvers and differentiable Python workflows. It is built upon NVIDIA AmgX~\cite{amgx2015}, a state-of-the-art library for GPU-accelerated sparse linear algebra that provides not only robust AMG algorithms but also a comprehensive suite of Krylov iterative methods. This integration brings scalable, differentiable sparse linear solvers and preconditioners to the JAX ecosystem.

Beyond simply exposing AmgX functionality in Python, JAX-AMG provides a unified interface deeply integrated with standard JAX workflows. It natively accepts JAX arrays and executes directly on GPU hardware, eliminating costly host-to-device memory transfers. Crucially, the library implements reverse-mode AD via discrete adjoint methods, seamlessly connecting the external sparse solver to JAX's transformation system through primitives such as \texttt{jax.grad}. It also fully supports JIT compilation via \texttt{jax.jit} for tracing and optimizing end-to-end simulation pipelines. To address the scaling demands of massive computational grids, the package enables distributed multi-GPU execution using GPU-aware MPI. Furthermore, JAX-AMG introduces intelligent caching utilities that store and reuse solver setup phases, such as the construction of the multigrid hierarchy. This significantly reduces computational overhead during the repeated solves characteristic of time-stepping, inverse problems, and PDE-constrained optimization.

Table~\ref{tab:comparison} summarizes how JAX-AMG addresses the gaps left by existing tools in the Python ecosystem. By unifying GPU acceleration, AMG, AD, JIT compilation, and MPI distribution in a single package, JAX-AMG resolves a critical bottleneck in modern differentiable scientific computing.

\begin{table}[!tbp]
\centering
\caption{Feature comparison of Python sparse solver packages relevant to JAX-based workflows.}
\label{tab:comparison}
\begin{tabular}{lccccc}
\toprule
 & GPU & AMG & AD & JIT & MPI \\
\midrule
\texttt{jax.scipy} \cite{jax2018github} (JAX-native) & \checkmark & --            & \checkmark & \checkmark & -- \\
PyAMG \cite{pyamg2023}       & --            & \checkmark & --            & --            & -- \\
pyamgx \cite{pyamgx}  (via AmgX)        & \checkmark    & \checkmark & --            & --            & -- \\
petsc4py \cite{petsc4py2011}  (via PETSc) & \checkmark & \checkmark & --         & --            & \checkmark \\
Lineax \cite{lineax2023}     & \checkmark    & --            & \checkmark & \checkmark & -- \\
AMJax \cite{amjax2026} (via PyAMG) & \checkmark$^\dagger$ & \checkmark & \checkmark & \checkmark & -- \\
\textbf{JAX-AMG}             & \checkmark    & \checkmark & \checkmark & \checkmark & \checkmark \\
\bottomrule
\end{tabular}

\vspace{0.3em}
\footnotesize{$^\dagger$ AMG hierarchy setup is CPU-side.}
\end{table}

\section{Software description}

JAX-AMG provides GPU-accelerated sparse linear solvers with full support for AD and JIT compilation in JAX. This section describes the software architecture and its principal functionalities.

\subsection{Software architecture}

\begin{figure}[!t]
    \centering
    \resizebox{0.8\columnwidth}{!}{

\begin{tikzpicture}[
    font=\small,
    every node/.style={inner sep=3pt, outer sep=0pt},
    >=Stealth,
    ibox/.style={
        rounded corners=3pt,
        line width=0.3pt,
        align=center,
        font=\footnotesize,
    },
    myarrow/.style={
        ->,
        >=Stealth,
        line width=0.5pt,
        draw=black!45,
    },
    layertitle/.style={
        font=\footnotesize\bfseries,
        text=black!70,
        anchor=north west,
    },
]

\def\oleft{-6.0}
\def\oright{6.0}



\node[ibox, draw=blue!50, fill=blue!15,
      minimum width=3.2cm, minimum height=0.42cm]
    at (-3.8, 0.59)
    {\scriptsize Sparse matrix};

\node[ibox, draw=blue!50, fill=blue!15,
      minimum width=3.2cm, minimum height=0.42cm]
    at (-3.8, 0.08)
    {\scriptsize Dense matrix};

\node[ibox, draw=blue!50, fill=blue!15,
      minimum width=3.2cm, minimum height=0.42cm]
    at (-3.8, -0.43)
    {\scriptsize Matrix-free operator};

\node[ibox, draw=blue!50, fill=blue!15,
      minimum width=3.2cm, minimum height=1.44cm, text width=2.8cm]
    (csr) at (0.0, 0.08)
    {{\scriptsize CSR}\\[-2pt]
     {\scriptsize sparse matrix}};

\node[ibox, draw=blue!50, fill=blue!15,
      minimum width=3.2cm, minimum height=1.44cm, text width=2.8cm]
    (solve) at (3.8, 0.08)
    {\textbf{\scriptsize \texttt{jaxamg.solve(A, b, config)}}};

\draw[myarrow] (-2.2, 0.08) -- (csr.west);
\draw[myarrow] (csr.east) -- (solve.west);

\begin{scope}[on background layer]
    \draw[rounded corners=6pt, draw=blue!40, fill=blue!5, line width=0.4pt]
        (\oleft, -0.94) rectangle (\oright, 1.50);
\end{scope}
\node[layertitle] at ([xshift=8pt]\oleft, 1.40) {Python API layer};

\draw[myarrow] (0, -0.94) -- node[right, font=\scriptsize, text=black!50, midway] {JAX} (0, -1.64);


\node[ibox, draw=teal!50, fill=teal!15,
      minimum width=3.2cm, minimum height=0.42cm]
    at (-3.8, -2.55)
    {\texttt{\scriptsize jax.jit}};

\node[ibox, draw=teal!50, fill=teal!15,
      minimum width=3.2cm, minimum height=0.42cm]
    at (-3.8, -3.06)
    {\texttt{\scriptsize jax.grad}};

\node[ibox, draw=teal!50, fill=teal!15,
      minimum width=3.2cm, minimum height=0.42cm]
    at (-3.8, -3.57)
    {\texttt{\scriptsize jax.vmap}};

\node[ibox, draw=teal!50, fill=teal!15,
      minimum width=3.2cm, minimum height=1.44cm, text width=2.8cm]
    (forward) at (0.0, -3.06)
    {\textbf{\footnotesize Forward}\\[1pt]
     {\scriptsize\texttt{jax.ffi.ffi\_call}}\\
     {\scriptsize $Ax = b$}};

\node[ibox, draw=teal!50, fill=teal!15,
      minimum width=3.2cm, minimum height=1.44cm, text width=2.8cm]
    (backward) at (3.8, -3.06)
    {\textbf{\footnotesize Backward}\\[1pt]
     {\scriptsize\texttt{jax.custom\_vjp}}\\
     {\scriptsize $A^\top \lambda = \partial\mathcal{L}/\partial x$}};

\draw[myarrow] (forward.east) -- (backward.west);

\begin{scope}[on background layer]
    \draw[rounded corners=6pt, draw=teal!40, fill=teal!5, line width=0.4pt]
        (\oleft, -4.08) rectangle (\oright, -1.64);
\end{scope}
\node[layertitle] at ([xshift=8pt]\oleft, -1.74) {JAX integration layer};

\draw[myarrow] (0, -4.08) -- node[right, font=\scriptsize, text=black!50, midway] {CUDA FFI} (0, -4.78);


\node[ibox, draw=orange!50, fill=orange!12,
      minimum width=2.55cm, minimum height=1.36cm, text width=2.25cm]
    at (-4.2, -6.16)
    {\textbf{\footnotesize AmgX}\\[1pt]
     {\scriptsize AMG}\\
     {\scriptsize Krylov methods}};

\node[ibox, draw=orange!50, fill=orange!12,
      minimum width=2.55cm, minimum height=1.36cm, text width=2.25cm]
    at (-1.4, -6.16)
    {\textbf{\footnotesize cuSPARSE}\\[1pt]
     {\scriptsize Single-device}\\
     {\scriptsize transpose $A^\top$}};

\node[ibox, draw=yellow!60!orange, fill=yellow!15,
      minimum width=2.55cm, minimum height=1.36cm, text width=2.25cm]
    at (1.4, -6.16)
    {\textbf{\footnotesize LRU Cache}\\[1pt]
     {\scriptsize Setup reuse}\\
     {\scriptsize Fixed sparsity}};

\node[ibox, draw=violet!50, fill=violet!12,
      minimum width=2.55cm, minimum height=1.36cm, text width=2.25cm]
    at (4.2, -6.16)
    {\textbf{\footnotesize MPI}\\[1pt]
     {\scriptsize GPU-aware}\\
     {\scriptsize Distributed $A^\top$}};

\begin{scope}[on background layer]
    \draw[rounded corners=6pt, draw=orange!40, fill=orange!5, line width=0.4pt]
        (\oleft, -7.14) rectangle (\oright, -4.78);
\end{scope}
\node[layertitle] at ([xshift=8pt]\oleft, -4.88) {C\texttt{++}/CUDA backend};

\draw[myarrow] (0, -7.14) -- (0, -7.84);

\node[ibox, draw=gray!50, fill=gray!10, minimum width=8cm] at (0, -8.14)
    {GPU (single or multi devices)};

\end{tikzpicture}}
    \caption{Software architecture of JAX-AMG.}
    \label{fig:architecture}
\end{figure}
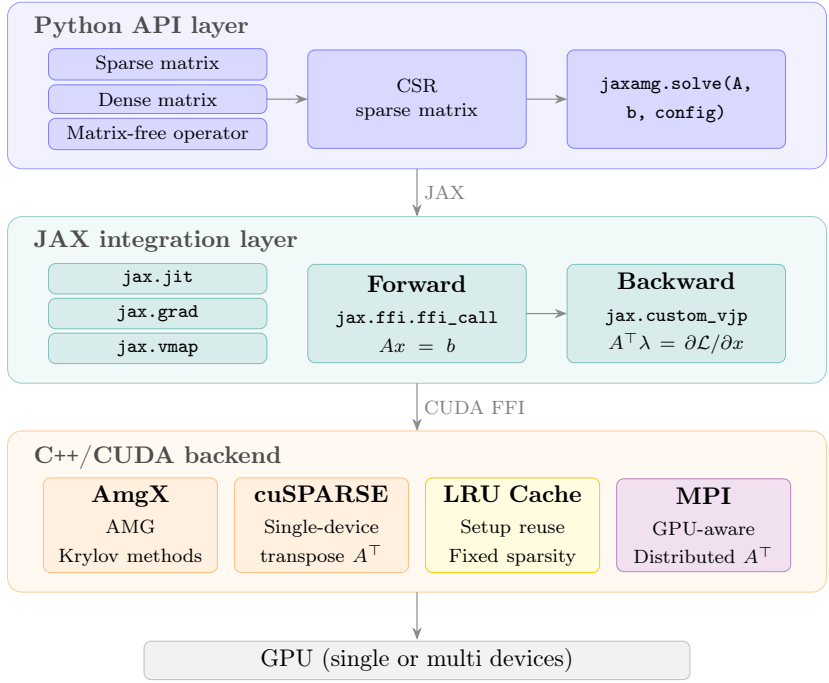

JAX-AMG is organized into three layers: a Python API layer, a JAX integration layer, and a native C++/CUDA backend, as illustrated in Fig.~\ref{fig:architecture}.

\textbf{Python API.}
The main public interface is a single \texttt{jaxamg.solve} entry point, modeled after the native sparse solvers in the \texttt{jax.scipy} module. It accepts a matrix $A$ and right-hand side $b$, along with an optional configuration dictionary for specifying solvers and preconditioners. Various input matrix formats are accepted, including JAX and SciPy sparse matrices, dense arrays, and callable linear operators. All inputs are normalized internally to the compressed sparse row (CSR) format. Solver configuration is specified via Python dictionaries, with the default being a BiCGSTAB solver with an AMG preconditioner.

\textbf{JAX integration.}
At the execution boundary, JAX-AMG registers CUDA foreign-function interface (FFI) targets with JAX's XLA compiler and dispatches solves through \texttt{jax.ffi.ffi\_call}, exposing the solver as a native JAX primitive that is fully compatible with \texttt{jax.jit} compilation. For AD, the library defines custom vector-Jacobian product (VJP) rules via \texttt{jax.custom\_vjp}. Given a forward solve $Ax = b$ and a scalar loss function $\mathcal{L}$, the backward pass computes the adjoint variable $\lambda$ via the implicit function theorem,
\begin{equation}
A^\top \lambda = \frac{\partial \mathcal{L}}{\partial x},
\label{eq:adjoint}
\end{equation}
from which the gradients follow as
\begin{equation}
\frac{\partial \mathcal{L}}{\partial A} = -\lambda x^\top, \qquad
\frac{\partial \mathcal{L}}{\partial b} = \lambda.
\end{equation}

For symmetric matrices, the transpose solve is skipped. For nonsymmetric systems, the backward pass forms $A^\top$ using cuSPARSE on a single GPU or an all-to-all exchange in MPI mode, and then solves the adjoint system through the same AmgX backend with the same solver and preconditioner configurations.

\textbf{Matrix-free operator support.}
JAX-AMG accepts matrix-free linear operators, specified as callables $x \mapsto Ax$, in addition to explicit sparse matrices. Because the AmgX backend requires explicit CSR data, the library materializes such an operator in two stages: it first recovers the operator's sparsity pattern, then assembles the nonzero values. Value assembly uses graph-colored probing: a coloring of the sparsity graph groups structurally independent columns, so that all nonzero values are recovered with a number of operator evaluations equal to the number of colors rather than the matrix dimension~\cite{curtis1974estimation,gebremedhin2005color}. This design targets sparse operators: the assembled CSR matrix must fit in memory, so operators that materialize as effectively large dense matrices may not be well suited to this approach.

The sparsity pattern itself is recovered by tracing: JAX-AMG interprets the operator's computation graph and propagates a sparse connectivity structure through each primitive, recovering the sparsity pattern in a single pass with no operator evaluations~\cite{hill2025sparser}. Following the similar strategy as \texttt{Sparse\allowbreak Connectivity\allowbreak Tracer.jl} in the Julia ecosystem, this approach scales to large operators at negligible memory. When an operator cannot be traced---for instance, one with opaque custom-call primitives or data-dependent indexing---the library falls back to probing with one-hot basis vectors. This fallback is correct for any operator but materializes dense probe batches on the GPU and incurs $\mathcal{O}(N)$ evaluations.

When called outside JIT, this materialization happens automatically and is cached on the operator object. When a matrix-free operator appears inside JIT-compiled code, for example in an optimization loop with a parameterized operator, the sparsity pattern and coloring are precomputed once via \texttt{cache\_coloring} and attached with \texttt{with\_cache}, because they fix the matrix's nonzero structure and color count, which JAX requires to be known at compile time. An illustrative example of this workflow is provided in Section~\ref{sec:opt-matfree-example}.

\textbf{C++/CUDA backend.}
The C++ layer manages AmgX resources (matrices, vectors, solvers, configurations) and exposes four FFI handlers covering single- and multi-GPU modes in both single and double precision. An LRU cache retains AmgX solver handles keyed by a hash of the full sparsity pattern, together with the matrix sparsity dimensions, precision, and solver configuration. On a cache hit, the matrix coefficients are updated in place and the solver setup is repeated against the new values, while the cached solver resources and matrix structure from the initial solve are reused. This is substantially cheaper than a cold setup and remains correct whenever the sparsity pattern is unchanged, as in typical PDE-constrained optimization. If the pattern itself changes (e.g., from topology-changing mesh adaptation), the cache key changes and a fresh setup is triggered automatically.

\textbf{Distributed execution.}
In MPI mode, each rank holds a local matrix partition and right-hand side. The same \texttt{jaxamg.solve} interface is used, with additional arguments for the MPI communicator and global problem size. When this communicator spans multiple nodes, the same setup scales across them without modification. JAX-AMG does not currently support automatic sharding of a global JAX sparse matrix/vector through \texttt{pjit} or \texttt{NamedSharding}; users instead supply the local partitions directly, aided by helper utilities such as \texttt{partition\_csr\_matrix} and \texttt{partition\_operator}. Communication is handled through mpi4jax \cite{mpi4jax}, which provides MPI collectives as JAX-traceable primitives, with support for GPU-aware MPI when available. In the backward pass, for nonsymmetric $A$, the distributed transpose is assembled by an all-to-all exchange through mpi4jax that routes each local nonzero $A_{ij}$ to the rank owning global row $j$ (its position in $A^{\top}$). AmgX then solves the resulting distributed $A^{\top}$ system. For matrix-value gradients in the distributed VJP, JAX-AMG exchanges only the remote solution entries referenced by each rank's local sparse rows, avoiding an all-gather of the full global solution vector.

\subsection{Software functionalities}

The principal functionalities of JAX-AMG are:
\begin{itemize}
    \item \textbf{GPU-accelerated sparse solves} using the full AmgX solver suite, including algebraic multigrid, Krylov subspace methods, and a variety of smoothers and preconditioners, all fully configurable via Python dictionaries compatible with AmgX's configuration format.
    \item \textbf{Reverse-mode automatic differentiation} through linear solves via adjoint methods (Eq.~\ref{eq:adjoint}), compatible with \texttt{jax.grad}. For symmetric matrices, the transpose  solve is automatically skipped.
    \item \textbf{JIT compilation} via \texttt{jax.jit} for traced, compiled execution of end-to-end pipelines.
    \item \textbf{Sequential \texttt{jax.vmap} compatibility} for applying the solver over a batch of right-hand sides. Because the matrix is shared across the batch, solver caching amortizes the setup cost over the batch instead of repeating it for each right-hand side.
    \item \textbf{Flexible input formats}, including JAX and SciPy sparse matrices, dense arrays, and matrix-free operators.
    \item \textbf{Automatic sparsity detection} for matrix-free operators, recovering the sparsity pattern by tracing the operator's computation graph in a single pass, with basis-vector probing as a fallback.
    \item \textbf{Distributed multi-GPU execution} via MPI, with support for GPU-aware MPI when available.
    \item \textbf{Preconditioner export} via \texttt{jaxamg.make\_preconditioner}, returning a callable compatible with native JAX sparse linear solvers in \texttt{jax.scipy} and third-party libraries such as Lineax.
    \item \textbf{Solver caching} at both the Python and C++ levels to amortize setup costs across repeated solves with fixed sparsity patterns.
    \item \textbf{Single and double precision} arithmetic, with automatic promotion of mixed-precision inputs.
\end{itemize}

Solver behavior is controlled through a Python dictionary interface. Users need only specify the parameters they wish to change; all other settings inherit sensible defaults. The solver returns diagnostic information including iteration count, final residual, and convergence status. Optional file-based output provides detailed AMG hierarchy statistics and per-iteration residual histories.

\section{Illustrative examples}

We present four examples that demonstrate the core capabilities of JAX-AMG, followed by performance comparisons against native JAX solvers and other AMG libraries. Additional examples are available in the repository's \texttt{demo} directory.

\subsection{GPU-accelerated sparse solve with AMG preconditioning}

The following example solves a 2D Poisson system on a $32 \times 32$ grid using BiCGSTAB with an AMG preconditioner:

\begin{lstlisting}[language=Python]
import jaxamg
from jaxamg.matrices import poisson_matrix, rhs_ones

n = 32
A = poisson_matrix(n)
b = rhs_ones(n*n)

x, info = jaxamg.solve(A, b, config={
    "solver": "PBICGSTAB",
    "preconditioner": {"solver": "AMG"},
    "tolerance": 1e-6,
})
\end{lstlisting}

This example converges in 6 iterations, whereas the same solver without preconditioning requires 37 iterations, illustrating the effectiveness of AMG preconditioning for this class of problem. For more severely ill-conditioned systems, convergence without preconditioning may not be achievable at all.

\subsection{Gradient-based optimization}
\label{sec:opt-example}

A key use case for JAX-AMG is embedding linear solves inside gradient-based optimization loops. The following example recovers the unknown diagonal value $d$ of a Toeplitz tridiagonal system such that the solution vector $x$ matches a target solution $x_{\mathrm{target}}$:

\begin{lstlisting}[language=Python]
import jax
import jax.numpy as jnp
import jaxamg
from jaxamg.matrices import tridiagonal_matrix, rhs_ones

n = 32
true_diag = 4.0
init_diag = 10.0
lr = 0.1
max_iters = 200
tol = 1e-3

A_true = tridiagonal_matrix(n, diagonal_value=true_diag)
b = rhs_ones(n)
x_target, _ = jaxamg.solve(A_true, b)

def loss(diag):
    A = tridiagonal_matrix(n, diagonal_value=diag)
    x, _ = jaxamg.solve(A, b)
    return jnp.sum((x - x_target) ** 2)

grad_fn = jax.jit(jax.value_and_grad(loss))
diag = init_diag
for step in range(max_iters):
    val, grad = grad_fn(diag)
    diag -= lr * grad
    if jnp.linalg.norm(grad) < tol:
        break
\end{lstlisting}

The gradient computation is JIT-compiled via \texttt{jax.jit}, with gradients obtained through \texttt{jax.value\_and\_grad}. In the backward pass, JAX-AMG automatically solves the adjoint system and computes gradients with respect to the matrix entries. Starting from an initial guess of $d = 10.0$, the optimizer recovers the true diagonal value $d = 4.0$ in fewer than 100 gradient steps under the specified learning rate and tolerance.

\subsection{Gradient-based optimization with matrix-free operators}
\label{sec:opt-matfree-example}

This example solves the same inverse problem as Section~\ref{sec:opt-example}, recovering the diagonal value $d$ of a Toeplitz tridiagonal system, but defines the operator $A$ as a matrix-free callable rather than an explicit sparse matrix. Because the operator depends on $d$ and changes at each optimization
step, the sparsity pattern and graph coloring are precomputed via
\texttt{cache\_coloring} before the optimization loop and attached via \texttt{with\_cache} inside the loss function, enabling JIT compilation. For this operator the sparsity pattern is recovered by tracing in a single pass with no operator evaluations.

\begin{lstlisting}[language=python]
import jax
import jax.numpy as jnp
import jaxamg
from jaxamg.matrices import rhs_ones

def A_operator(diag):
    def matvec(x):
        y = -jnp.roll(x, 1) + diag * x - jnp.roll(x, -1)
        y = y.at[0].set(diag * x[0] - x[1])
        y = y.at[-1].set(-x[-2] + diag * x[-1])
        return y
    return matvec

n = 32
true_diag = 4.0
init_diag = 10.0
lr = 0.1
max_iters = 200
tol = 1e-3

A_true = A_operator(true_diag)
b = rhs_ones(n)
x_target, _ = jaxamg.solve(A_true, b)

diag = init_diag
coloring_cache = jaxamg.cache_coloring(A_operator(init_diag), shape=n)

def loss(diag):
    A = jaxamg.with_cache(A_operator(diag), coloring=coloring_cache)
    x, _ = jaxamg.solve(A, b)
    return jnp.sum((x - x_target) ** 2)

grad_fn = jax.jit(jax.value_and_grad(loss))

for step in range(max_iters):
    val, grad = grad_fn(diag)
    diag -= lr * grad
    if jnp.linalg.norm(grad) < tol:
        break
\end{lstlisting}

The matrix-free formulation produces identical results to the explicit matrix case in Section~\ref{sec:opt-example}, recovering $d = 4.0$ from an initial guess of $d = 10.0$ within the same number of gradient steps.

\subsection{Distributed multi-GPU solve}

JAX-AMG supports MPI-distributed solves with minimal changes to the single-GPU interface. The following example partitions a 2D Poisson problem across MPI ranks and solves it in parallel:

\begin{lstlisting}[language=Python]
from mpi4py import MPI
import jaxamg
from jaxamg.matrices import poisson_matrix_distributed, rhs_linear
from jaxamg.mpi_utils import partition_vector, gather_vector

comm = MPI.COMM_WORLD
rank, nranks = comm.Get_rank(), comm.Get_size()

n = 32
A_local, row_start, row_end = poisson_matrix_distributed(
    n, n, rank, nranks)
b_local, _, _ = partition_vector(rhs_linear(n*n), rank, nranks)
x_local, info = jaxamg.solve(
    A_local, b_local, comm=comm, nglobal=n*n,
    partition_info=(row_start, row_end),
    config={
        "solver": "PBICGSTAB",
        "preconditioner": {"solver": "MULTICOLOR_DILU"},
        "communicator": "MPI_DIRECT"})
x_global = gather_vector(x_local, comm, root=0)
jaxamg.finalize()
\end{lstlisting}

\subsection{Performance comparison}
\label{sec:performance}
\begin{figure}[!tbp]
    \centering
    \includegraphics[width=\linewidth]{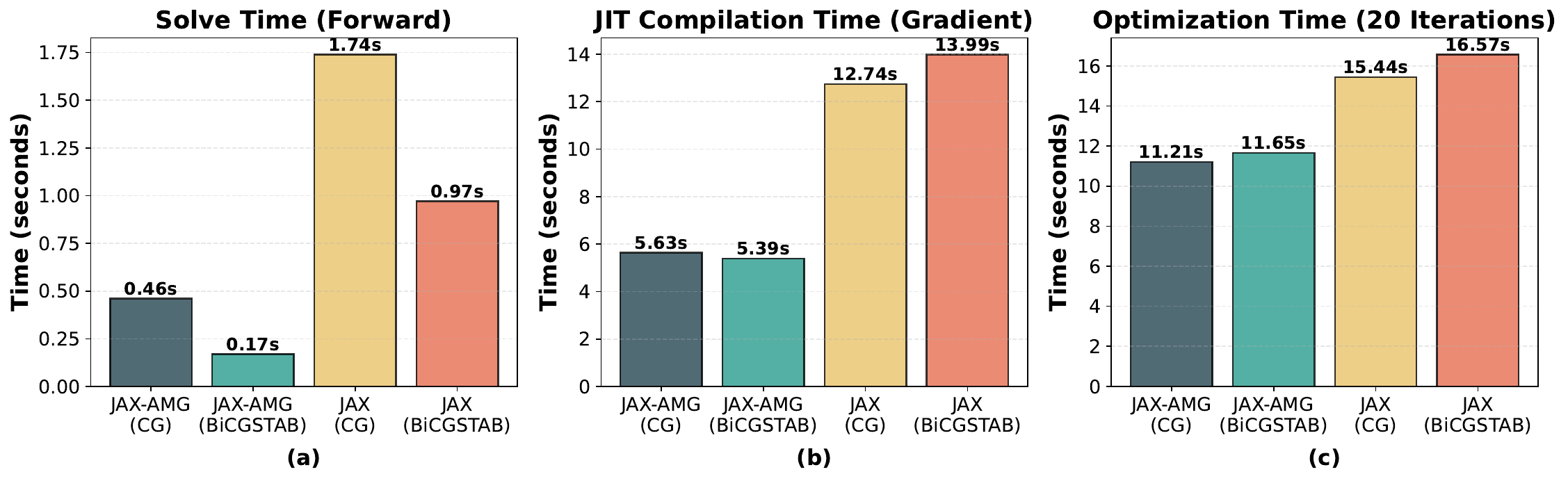}
    \caption{Performance comparison between JAX-AMG and native JAX solvers (CG and BiCGSTAB) on a tridiagonal system with $n = 10^7$ unknowns. (a)~Forward solve wall-clock time. (b)~JIT compilation time for gradient calculation. (c)~Total optimization time for 20 iterations. All solvers use the same unpreconditioned configuration. Benchmarked on an NVIDIA L40 GPU.}
    \label{fig:performance}
\end{figure}

\textbf{Comparison with native JAX solvers.} To evaluate the performance of JAX-AMG, we first compare it against the native JAX CG and BiCGSTAB sparse linear solvers on a tridiagonal system of dimension $n = 10^7$. To isolate the effect of the solver backend from that of preconditioning or other algorithmic differences, we deliberately choose a well-conditioned system that the native JAX solvers can handle without preconditioning, and configure JAX-AMG to use the same unpreconditioned CG and BiCGSTAB methods. Fig.~\ref{fig:performance}a compares wall-clock time for a single forward solve, Fig.~\ref{fig:performance}b reports JIT compilation time for the gradient computation, and Fig.~\ref{fig:performance}c shows the total optimization time over 20 iterations. In all three cases, JAX-AMG outperforms native JAX solvers, demonstrating the performance advantage of the GPU-optimized AmgX backend.

\begin{figure}[!tbp]
    \centering
    \includegraphics[width=\linewidth]{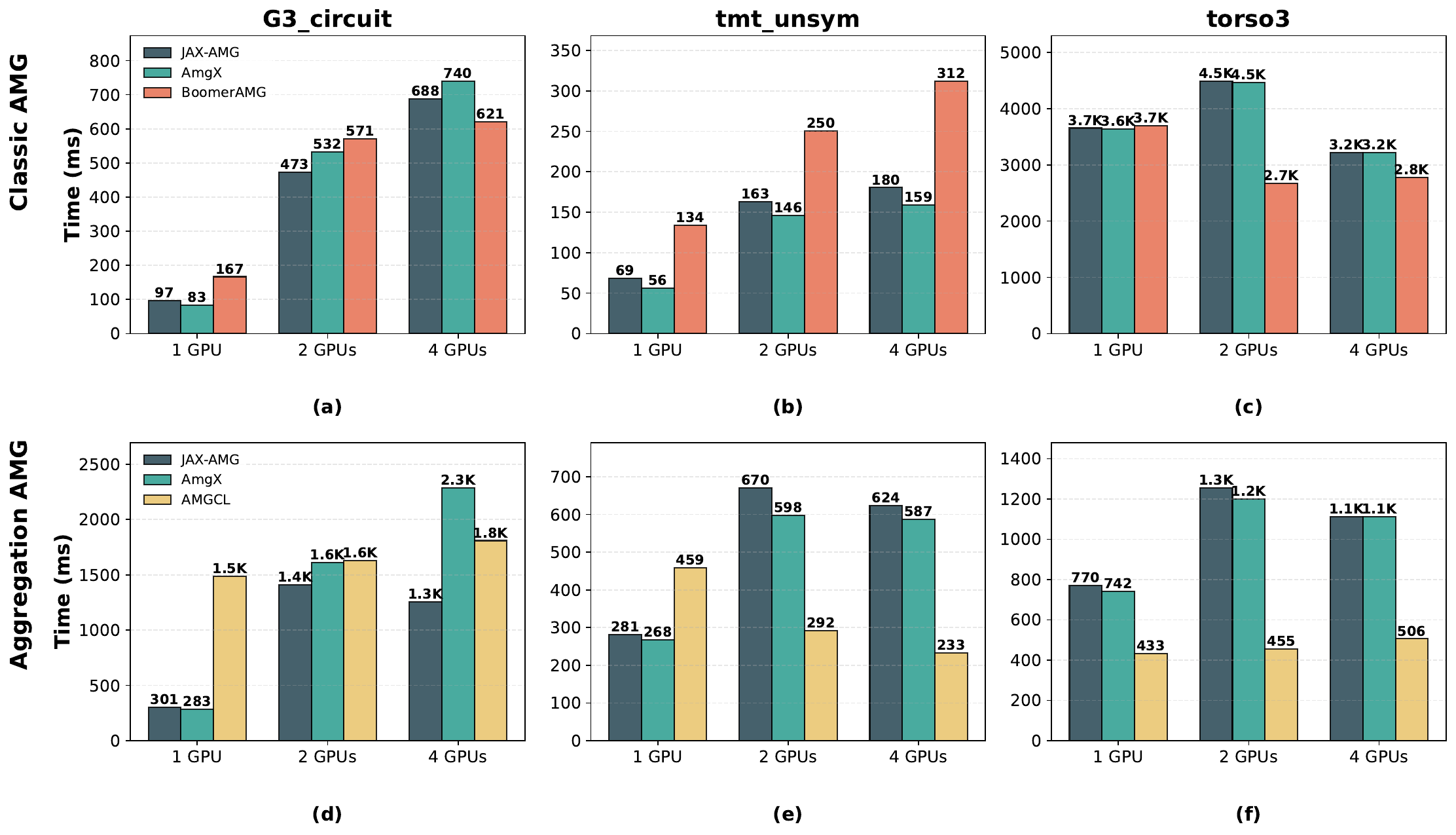}
    \caption{Performance comparison of classical and aggregation AMG solve times for three sparse matrices across 1, 2, and 4 GPUs. Panels (a–c) show classical AMG results (JAX-AMG, AmgX, and BoomerAMG) for matrices \texttt{G3\_circuit}, \texttt{tmt\_unsym}, and \texttt{torso3}, respectively. Panels (d–f) show aggregation AMG results (JAX-AMG, AmgX, and AMGCL) for the same three matrices. Within each panel, the $x$-axis represents GPU count (1, 2, or 4 GPUs), and bars show performance for each solver. Reported timings measure a warm resetup-and-solve step on NVIDIA L40 GPUs.}
    \label{fig:performance2}
\end{figure}

\textbf{Comparison with established AMG libraries.} We next benchmark JAX-AMG against the native AmgX backend and two widely used AMG libraries (BoomerAMG and AMGCL) on three large, structurally distinct sparse systems from the SuiteSparse Matrix Collection~\cite{davis2011university}: \texttt{G3\allowbreak \_circuit} (circuit simulation; structurally and value-symmetric), \texttt{tmt\_unsym} (electromagnetic simulation; structurally symmetric but value-nonsymmetric), and \texttt{torso3} (electrophysiological modeling; structurally nonsymmetric). These systems span different application domains and sparsity patterns. Each system is solved with a Krylov iterative method preconditioned by classical Ruge--St\"uben and aggregation-based AMG on 1, 2, and 4 GPUs using the same configurations. We report the cost of a warm resetup-and-solve, because this is the operation typically repeated within optimization and implicit time-stepping loops, where the matrix entries change but the sparsity pattern remains fixed, and therefore dominates the end-to-end cost. Figure~\ref{fig:performance2} presents the six resulting configurations (three matrices $\times$ classical/aggregation AMG), comparing native AmgX and JAX-AMG across the three GPU counts, together with BoomerAMG~\cite{yang2002boomeramg} for the classical AMG panels and AMGCL~\cite{demidov2019amgcl} for the aggregation AMG panels (BoomerAMG is classical-only, whereas AMGCL's MPI path is aggregation-only). Because AMG libraries expose different configurable options, we matched the solver settings as closely as possible when exact equivalence was not available. These benchmarks primarily compare solver implementations, and the GPU-count trends should be interpreted with care because distributed resetup and communication overheads can produce non-monotonic timings for these matrix sizes. Across all cases, JAX-AMG closely tracks native AmgX, confirming that the JAX FFI introduces negligible overhead on the repeated solve path. On multiple GPUs, JAX-AMG is sometimes even slightly faster than the native backend. Since both backends issue identical AmgX operations, this is not an algorithmic advantage but a runtime effect, likely related to how the JAX/XLA runtime schedules host work around MPI progress and communication overlap during distributed resetup. Relative to other established libraries, AmgX/JAX-AMG is competitive, although no AMG library is consistently superior. We note that AmgX performs resetup entirely on the GPU, whereas BoomerAMG offers no lightweight resetup and must rebuild its hierarchy from scratch, and AMGCL performs its rebuild on the host CPU.

\begin{figure}[!tbp]
    \centering
    \includegraphics[width=\linewidth]{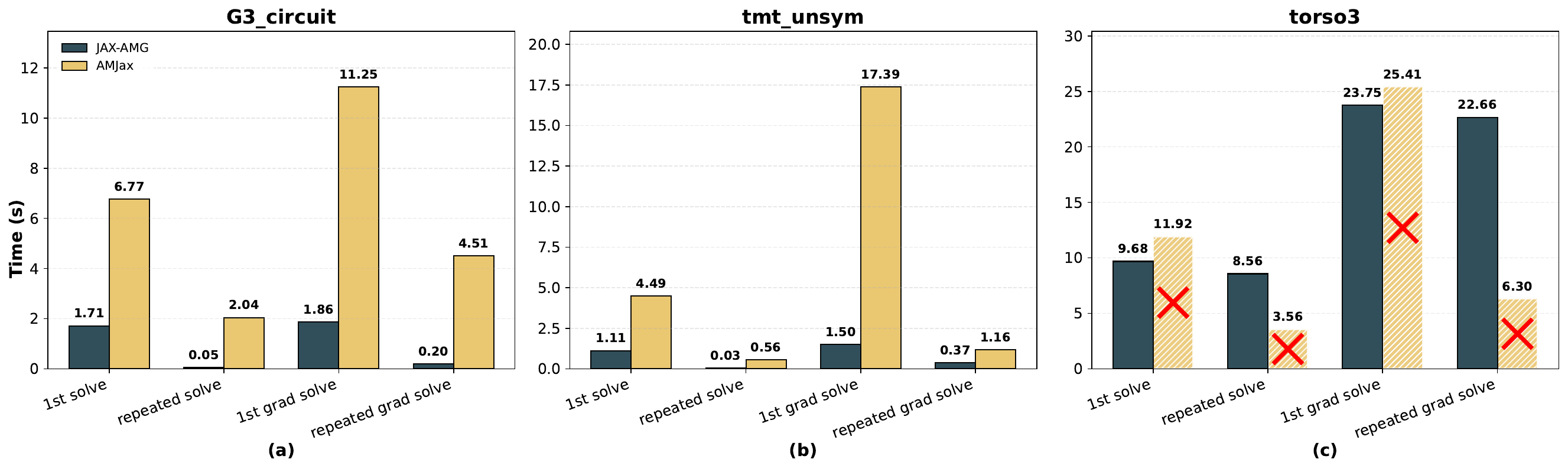}
    \caption{Performance comparison of JAX-AMG and AMJax on three benchmark matrices (\texttt{G3\_circuit}, \texttt{tmt\_unsym}, and \texttt{torso3}). Each panel reports wall-clock times for the first forward solve, subsequent forward solve, first gradient solve, and subsequent gradient solve. Hatched bars overlaid with a red cross marker denote AMJax runs that failed to converge. Benchmarked on an NVIDIA L40 GPU with 48~GB of memory.}
    \label{fig:performance3}
\end{figure}

\textbf{Comparison with AMJax.} We next compare JAX-AMG with AMJax \cite{amjax2026}, a JAX-oriented AMG solver built on PyAMG. In this comparison, AMJax is used as a preconditioner for JAX-native Krylov solvers, with the solver and preconditioner configurations matched as closely as the two libraries allow. The comparison uses the same three sparse matrices considered above. Since AMJax does not provide an AmgX-style resetup operation, we compare fixed-hierarchy repeated solves and run JAX-AMG with \texttt{reuse\_setup=True} to reuse the existing hierarchy without resetup. The experiments are limited to a single GPU because AMJax does not support distributed execution. The results are shown in Fig.~\ref{fig:performance3}. On \texttt{G3\_circuit} and \texttt{tmt\_unsym}, AMJax converges but is substantially slower than JAX-AMG for both forward and gradient computations. On the most challenging case, \texttt{torso3}, JAX-AMG still converges, whereas AMJax diverges under the closest matched configuration, and additional tuning attempts also failed to produce valid forward or gradient solves. While AMJax offers a more portable JAX-based workflow, JAX-AMG achieves stronger robustness and performance for the large-scale GPU AMG workloads considered here, primarily due to its efficient AmgX backend and richer solver configurations.

\begin{figure}[!tbp]
    \centering
    \includegraphics[width=1.0\textwidth]{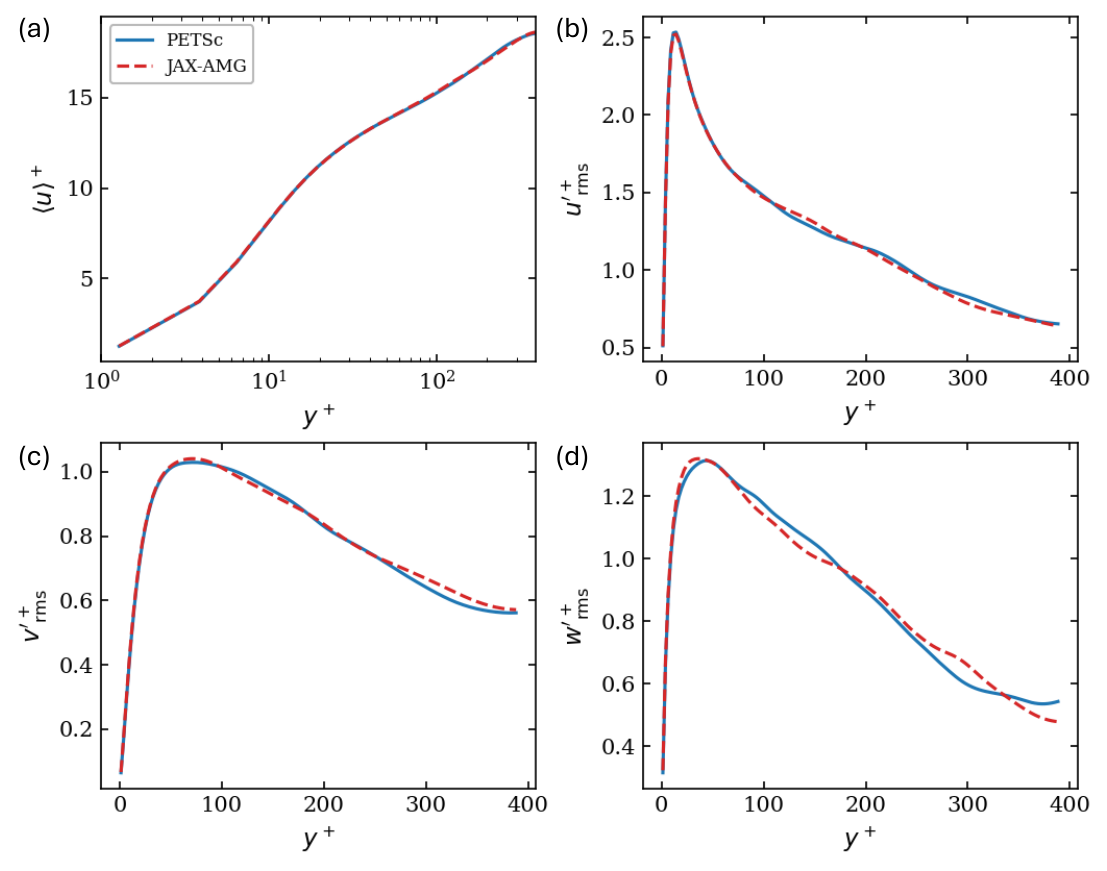}
    \caption{Velocity statistics from Diff-FlowFSI using JAX-AMG and PETSc as the Poisson solver for turbulent channel flow at $Re_{\tau}=390$: (a)~mean streamwise velocity; (b--d)~root-mean-square of streamwise, wall-normal, and spanwise velocity, respectively.}
    \label{fig:velocity_statistics}
\end{figure}

\textbf{Comparison with PETSc for CFD application.} Finally, we test JAX-AMG on a practical problem where preconditioning is essential, integrating it as the pressure Poisson solver in Diff-FlowFSI~\cite{fan2026diff}, our in-house differentiable incompressible CFD solver, to simulate turbulent channel flow at friction Reynolds number $Re_{\tau}=390$, where unpreconditioned BiCGSTAB fails to converge. The computational domain is $L_x \times L_y \times L_z = 2\pi \times 1 \times \pi$, discretized on a $100 \times 260 \times 256$ grid with wall-normal stretching. We compare JAX-AMG against PETSc~\cite{petsc1997,petsc4py2011}, both configured with BiCGSTAB and multigrid preconditioning. In this stretched-grid finite-volume case, the stored pressure Poisson matrix is nonsymmetric, so BiCGSTAB is used. Compatibility and zero-mean constraints are enforced for the periodic/Neumann pressure setup to remove the constant null mode. As shown in Table~\ref{tab:petsc_comparison}, JAX-AMG requires fewer iterations and lower per-GPU memory than PETSc under the same convergence tolerance, at the cost of a moderate increase in solve time. On one and two GPUs, JAX-AMG converges in 5--6 iterations compared with 15 for PETSc, while reducing per-GPU memory by about 29\% and 41\%, respectively. The two-GPU timings correspond to fixed-problem-size distributed runs, where communication and distributed-solver overheads can offset the reduced local workload. Figure~\ref{fig:velocity_statistics} confirms that the resulting turbulence statistics are in close agreement. Small discrepancies in second-order statistics between the two results are expected because the solvers use different AMG implementations and configuration options. Crucially, unlike PETSc, JAX-AMG retains full compatibility with JAX's AD and JIT compilation, enabling gradient-based optimization through the same solver without breaking the differentiable pipeline.

\begin{table}
    \centering
    \begin{tabular}{l c c c c}
    \toprule
        \multirow{2}{*}{Item} & \multicolumn{2}{c}{Single GPU} & \multicolumn{2}{c}{2 GPUs} \\
        \cmidrule(lr){2-3} \cmidrule(lr){4-5}
         & PETSc & JAX-AMG & PETSc & JAX-AMG \\
        \midrule
        Time (s) & 2.90 & 3.88 & 2.80 & 4.25 \\
        Memory (MiB)/GPU & 11776 & 8348 & 8758 & 5130 \\
        Convergence iteration & 15 & 5 & 15 & 6 \\
        Convergence error & $10^{-5}$ & $10^{-5}$ & $10^{-5}$ & $10^{-5}$ \\
    \bottomrule
    \end{tabular}
    \caption{Computational cost per Poisson solve comparing JAX-AMG and PETSc in the Diff-FlowFSI solver for incompressible turbulent channel flow at $Re_{\tau}=390$. Benchmarked on NVIDIA RTX A6000 GPUs.}
    \label{tab:petsc_comparison}
\end{table}

\section{Impact}

JAX-AMG fills a specific gap in the scientific computing toolchain: it is, to our knowledge, the first package to combine GPU-accelerated AMG with AD, JIT compilation, and distributed execution within the JAX ecosystem. A growing number of JAX-based differentiable simulation frameworks have been developed across various domains~\cite{newbury2024review}, including finite element methods~\cite{xue2023jax}, CFD~\cite{fan2026diff,Kochkov2021-ML-CFD}, and nanoscale heat transfer~\cite{shang2025jax}, among others. Many of these frameworks rely on sparse linear solves as a core computational kernel, yet the solver options currently available in JAX are limited: native solvers lack robust preconditioning, while external GPU solver libraries lack differentiability and JIT compatibility. JAX-AMG resolves this tension, and its impact extends across several areas.

The most direct impact is on PDE-constrained optimization and inverse problems. These workflows require differentiating through repeated sparse solves, often involving ill-conditioned systems where unpreconditioned Krylov methods either converge slowly or fail entirely. Because JAX-AMG provides AMG-preconditioned solves with gradient support, problems such as design optimization, optimal control, and spatially varying parameter inference can now be tackled end-to-end within JAX. Prior to JAX-AMG, researchers in the JAX ecosystem faced a choice between native solvers that lacked robust preconditioning and external solvers that broke the differentiable pipeline. JAX-AMG eliminates this trade-off.

A second area of impact is large-scale differentiable simulation. 
The MPI-distributed backend partitions problems across multiple GPUs while retaining full gradient support, extending differentiable  optimization to problem scales that exceed single-GPU memory. This is particularly relevant for high-resolution 3D PDE systems---such  as turbulent flow or full-domain cardiac hemodynamics simulations---where single-GPU memory constraints would otherwise preclude gradient-based  inversion or optimal control.

Beyond enabling new research, JAX-AMG improves the efficiency and flexibility of existing JAX-based workflows. Even for well-conditioned systems where preconditioning is unnecessary, the GPU-optimized AmgX backend achieves considerable speedups over native JAX solvers as demonstrated in Section~\ref{sec:performance}. The \texttt{jaxamg.\allowbreak{}make\_preconditioner} function allows JAX-AMG to serve as a preconditioning backend for native JAX or other third-party iterative solvers, bringing AMG preconditioning to solver pipelines that previously had access only to simple preconditioners. The unified Python interface, covering solver configuration, gradient computation, JIT compilation, and MPI distribution, reduces engineering effort and improves reproducibility compared to workflows that stitch together tools across multiple languages and runtimes.

JAX-AMG is publicly available at \url{https://github.com/jx-wang-s-group/JAX-AMG}  under the Apache 2.0 license. It has already been integrated into  Diff-FlowFSI~\cite{fan2026diff}, a differentiable CFD platform for  turbulent flow and fluid-structure interaction, where it serves as the pressure Poisson solver enabling gradient-based optimization through high-Reynolds-number simulations. It has similarly been adopted in JAX-BTE~\cite{shang2025jax}, a  differentiable solver for nanoscale phonon transport, demonstrating 
JAX-AMG's applicability across diverse PDE domains. These integrations validate JAX-AMG as a practical computational 
substrate for differentiable simulation in both fluid mechanics and 
thermal transport.

\section{Conclusions}

JAX-AMG bridges the gap between high-performance GPU sparse solvers and the JAX differentiable programming ecosystem. By wrapping the full NVIDIA AmgX solver suite, including AMG and Krylov methods, as a native JAX primitive, it brings robust GPU-accelerated sparse linear algebra into differentiable workflows with support for reverse-mode AD, JIT compilation, \texttt{jax.vmap} compatibility, and MPI-distributed execution. These capabilities make JAX-AMG a practical foundation for the growing ecosystem of JAX-based differentiable simulation frameworks that rely on sparse linear solves as a core computational kernel.

\section*{Acknowledgements}

The authors gratefully acknowledge funding from the U.S. Office of Naval Research under award number N00014-23-1-2071 and the U.S. National Institutes of Health under award number 1R01HL177814.

Research and computational support was provided by the NVIDIA Academic Grant Program. This work utilized NVIDIA GPUs (L40, RTX A6000) and software including the NVIDIA AmgX library, CUDA Toolkit, and cuSPARSE, 
which were essential for developing and benchmarking the GPU-accelerated sparse linear solver library JAX-AMG.





\bibliographystyle{elsarticle-num}
\bibliography{jaxamg}

\end{document}